\newcommand\Eq[1]{Eq.~(\ref{#1})}

\newcommand\Fig[1]{Fig.~\ref{#1}}
\newcommand\dm{$\rho$}

\newcommand\tHam{${\cal H}$}

\newcommand\bra[1]{\mbox{$\langle\,{#1}\,|$}}
\newcommand\ket[1]{\mbox{$|\,{#1}\,\rangle$}}

\newcommand\Id{\mbox{$\openone$}}
\newcommand\vol[1]{\textbf{#1}}
\newcommand\InnerProduct[2]{\mbox{$\langle\,{#1}\,|\,{#2}\,\rangle $}}

\documentclass[aps,pra,twocolumn, groupedaddress,amsmath,amssymb]{revtex4}

\usepackage{graphicx}
\usepackage{bm}

\begin{document}

% Use the \preprint command to place your local institutional report
% number in the upper righthand corner of the title page in preprint mode.
% Multiple \preprint commands are allowed.
% Use the 'preprintnumbers' class option to override journal defaults
% to display numbers if necessary

%\preprint{}

%Title of paper
\title{An exact mapping between the states of arbitrary N-level quantum systems \\
and the positions of classical coupled oscillators}

\author{Thomas E.~Skinner}
\email{thomas.skinner@wright.edu}  
\affiliation{Physics Department, Wright State University, Dayton, OH 45435}

\date{\today}

\begin{abstract}
The dynamics of states representing arbitrary N-level quantum systems, including dissipative systems, can be modeled exactly by the dynamics of classical coupled  oscillators. There is a direct one-to-one correspondence between the quantum states and the positions of the oscillators. Quantum coherence, expectation values, and measurement probabilities for system observables can therefore be realized from the corresponding classical states. The time evolution of an N-level system is represented as the rotation of a real state vector in hyperspace, as previously known for density matrix states but generalized here to Schr\"odinger states.  A single rotor in $n$-dimensions is then mapped directly to $n$ oscillators in one physical dimension.  
The number of oscillators needed to represent N-level systems scales linearly with $N$ for Schr\"odinger states, in contrast to $N^2$ for the density matrix formalism.
Although the well-known equivalence (SU(2), SO(3) homomorphism) of 2-level quantum dynamics to a rotation in real, physical space cannot be generalized to arbitrary N-level systems, representing quantum dynamics by a system of coupled harmonic oscillators in one physical dimension \textit{is} general for any N.
Values for the classical coupling constants are readily obtained from the system Hamiltonian, allowing construction of classical mechanical systems that can provide visual insight into the dynamics of abstract quantum systems as well as a metric for characterizing the interface between quantum and classical mechanics.
\end{abstract}

% insert suggested PACS numbers in braces on next line

% insert suggested keywords - APS authors don't need to do this
%\keywords{}

%\maketitle must follow title, authors, abstract, \pacs, and \keywords
\maketitle

\section{Introduction.}

The density matrix formalism \cite{Tolman, vonNeumann, Fano} provides a straightforward procedure for predicting quantum dynamics.  
At any given time, the density matrix provides a complete statistical characterization of the system in terms of the mean values of measurable states.  
It includes both the quantum uncertainty in predicting the results of single measurements on pure states (even though such states represent maximal possible information for the system) and the classical uncertainty in measurements on mixed states of less than maximal information.

Although the theory needs no supporting visual model for its application, the Liouville-von Neumann equation governing the time evolution of the density matrix provides little physical insight into system dynamics.
There has therefore been considerable effort towards representing, where possible, quantum systems using more intuitive classical models.  Of particular influence and importance is the classical representation for quantum two-level systems \cite{Feynman}, sometimes referred to as the Feynman--Vernon--Hellwarth (FVH) Theorem.  The behavior of any quantum mechanical two-level system can be modeled by classical torque equations, providing a one-to-one correspondence between the time evolution of the system and the dynamics of a spinning top in a constant gravitational field or a magnetic moment in a magnetic field. 

Work by Fano \cite{Fano} published concurrently with the FVH result also provides geometrical interpretation of spin dynamics for more complex quantum systems.  The density matrix for an N-level system is represented as an expansion in Hermitian operators, resulting in a vector with real components.  The time development of this generalized Bloch vector is a real rotation in a hyperspace of $(N^2 - 1)$ dimensions.  Constants of the motion can be derived \cite{Hioe, Elgin} that constrain the system's dynamics and provide physical insight.  However, the states of the system as given by the components of this vector (also referred to as a coherence vector \cite{Hioe} and, more recently, a Stoke's tensor \cite{Jaeger}), do not evolve in a physical space amenable to visualization, with its attendant advantages, except for the case $N=2$.

Thus, no completely general mapping has been realized that yields a one-to-one correspondence, similar to the FVH result, between the states of a quantum-mechanical N-level system and classical dynamical variables, providing the possibility for direct mechanical insight into the dynamics of abstract quantum systems. Analogies between quantum and classical systems have been noted \cite{Dirac, Strocchi, Frenkel, Anderson, Scott, Sullivan, Spreeuw, Frank, Jolk, Alzar, Novotny, Kovaleva, Elze} almost from the beginning.  But exact equivalence between the quantum and classical equations of motion has been obtained only for certain limiting conditions \cite{McKibben, Maris, Hemmer, Marx, Leroy, Shore, Briggs2012a, Eisfeld} such as weak perturbations of the system (weak coupling limit) and the aforementioned 2-level systems.  

Recently, the possibility of exact representations of N-level quantum systems in terms of classical coupled oscillators, with no restriction to weak coupling, was demonstrated \cite{Briggs2012b}.  However, the formalism is limited to real, invertible Hamiltonians applied only to pure states.  It is insufficiently general to represent statistical mixtures and density matrix evolution classically, as well as open, dissipative systems. 

In the present work, a very simple approach is presented for mapping an arbitrary N-level quantum system to a system of coupled harmonic oscillators, available for over a decade. It both complements and augments the results presented in \cite{Briggs2012b} and is made more relevant by that study.  The salient features representing the dynamics of N-level systems as rotations in Liouville space are reviewed first.  The desired one-to-one correspondence between the states of the quantum system, represented as a density matrix, and classical dynamical variables is provided by a mapping representing harmonic oscillators.  The quantum states, either pure or mixed, are represented exactly by the time-dependent displacements of classical coupled oscillators. 

Hilbert space rotations are reviewed next to generalize this option for representing quantum spin dynamics classically.  Results inspired by \cite{Briggs2012b} are presented with no restriction to real, invertible Hamiltonians.  An exact mapping of Schr\"odinger states to the physical displacements of coupled oscillators is provided, in contrast to the displacement and velocity originally required.  This approach to representing spin dynamics is then extended to mixed states.  Whereas $N^2-1$ oscillators are needed most generally to represent density matrix dynamics for an N-level system classically, Schr\"odinger states require at most $2N$ oscillators.
% , including a method for reducing the number of oscillators needed using the minimum set of pure states representing the density matrix [ref]???.  

Open (dissipative) N-level systems are considered next, showing they also can be exactly and simply modeled as classical coupled oscillators.  
The present treatment reveals the necessity for negative couplings in closed systems as well as antisymmetric couplings in open systems.

The paper closes with illustrative examples of the quantum-classical mapping.  

\section{Time Evolution of Quantum N-Level Systems}
\label{sec:Time Evolution}

A brief synopsis of formalisms for representing the dynamics of N-level systems is presented.  The standard Liouville space and Hilbert space representations are considered first, enabling simple generalizations that lend themselves to a classical interpretation.  In all cases, the time evolution of the system can be reduced to the form
%--------------------------------------------------------------
     \begin{equation}
\Phi(t) = U(t)\, \Phi(0).
\label{TimeEvolutionEq}
     \end{equation}
%--------------------------------------------------------------
The representation chosen determines the particular forms for $\Phi$ and the propagator $U(t)$.  For notational convenience and interchangeability of energy and frequency units, $\hbar$ is set equal to 1 in what follows.

\subsection{Liouville equation}
\label{subsec:Liouville}

The Liouville-von Neumann equation for the time evolution of a density matrix \dm\ governed by system Hamiltonian $H$ is 
%--------------------------------------------------------------
     \begin{equation}
\dot\rho  = -i\,[\,H,\rho\,],
\label{Liouville}
     \end{equation}
%--------------------------------------------------------------
with formal solution
%--------------------------------------------------------------
     \begin{eqnarray}
\rho(t)  &=& e^{-i\,H t}\,\rho(0)\, e^{i\,H t} \nonumber \\
         &=& U \rho(0)\, U^\dag,
\label{SolLiouville}
     \end{eqnarray}
%--------------------------------------------------------------
which defines $U(t) = e^{-i\,H\, t}$.

The time evolution can be related to a rotation by first expanding \dm\ in terms of a complete set of basis operators \cite{Fano}.  Orthogonal bases are particularly convenient and are typically normalized for further convenience. Denoting the basis elements as $\bm{\hat e}_i$ for state $i$ and requiring only that the basis be orthonormal gives
%--------------------------------------------------------------
     \begin{equation}
\langle\,\bm{\hat e}_i\,|\,\bm{\hat e}_j\,\rangle 
    = \text{Tr}\,(\bm{\hat e}_i^\dag\,\bm{\hat e}_j) = \delta_{ij},
\label{orthn}
      \end{equation}
%--------------------------------------------------------------
where the inner product for the vector space comprised of matrices is given by the  operator Tr, which returns the trace (sum of diagonal elements) of its argument, and $\dag$ denotes the operation of Hermitian conjugation.  In lieu of explicitly normalizing the $\bm{\hat e}_i$, the inner product can be defined with the appropriate factor multiplying Tr.
Summing over repeated indices gives \dm\ as
%--------------------------------------------------------------
     \begin{equation}
\bm{\rho} = r_j \bm{\hat e}_j 
\label{rho(r_j)}
      \end{equation}
%--------------------------------------------------------------
where the coefficients in the expansion are the projection onto the basis states.  Each $r_j$ in \Eq{rho(r_j)},
%--------------------------------------------------------------
	\begin{equation}
r_j = \langle\,\bm{\hat e}_j\,|\,\bm{\rho}\,\rangle 
    = \text{Tr}\,(\bm{\hat e}_j^\dag\,\bm{\rho})\, ,
\label{r}
        \end{equation}
%--------------------------------------------------------------
is thus the expectation value of the quantum state $\bm{\hat e}_j$.

Then
%--------------------------------------------------------------
	\begin{eqnarray}
\dot r_i = \text{Tr}(\bm{\hat e}_i^\dag\,\bm{\dot\rho}) 
  &=& -i\,\text{Tr}(\bm{\hat e}_i^\dag\,[\,H,\bm{\rho}\,]) \nonumber \\
  &=& -i\,\text{Tr}
      (\bm{\hat e}_i^\dag\,[\,H,\bm{\hat e}_j\,])\,r_j
            \nonumber \\
  &=& \Omega_{i j} r_j.
\label{rdot}
        \end{eqnarray}
%--------------------------------------------------------------
Expanding the commutator, using $\text{Tr}(AB) = \text{Tr}(BA) = [\text{Tr}(AB)^\dag]^*$ and
$[\,\bm{\hat e}_i^\dag\, ,\bm{\hat e}_j\,]^\dag = [\,\bm{\hat e}_j^\dag\, ,\bm{\hat e}_i\,]$ 
gives
%--------------------------------------------------------------
	\begin{eqnarray}
\Omega_{i j} &=&  -i\,\text{Tr}
      (\bm{\hat e}_i^\dag\,[\,H,\bm{\hat e}_j\,]) \nonumber \\
     &=& i\,\text{Tr}
      (\,[\,\bm{\hat e}_i^\dag\, ,\bm{\hat e}_j\,]\,H\,) \nonumber \\
     &=& -\{\,i\,\text{Tr}
      (\,[\,\bm{\hat e}_j^\dag\,,\bm{\hat e}_i\,]\,H\,)\}^* \nonumber \\
     &=& -\Omega_{j i}^*
\label{Omega_ij}
        \end{eqnarray}
%--------------------------------------------------------------
in terms of its complex conjugate elements, denoted by *.
Thus, $\Omega = -\Omega^\dag$ is antihermitian and can be diagonalized.
The evolution of the density matrix is given by
%--------------------------------------------------------------
	\begin{equation}
\bm{\dot r} = \Omega\, \bm{r},
\label{BlochEqMat}
	\end{equation}
%--------------------------------------------------------------
with solution
%--------------------------------------------------------------
	\begin{equation}
\bm{r}(t) = e^{\Omega t}\bm{r}(0).
\label{BlochSol}
   \end{equation}
%--------------------------------------------------------------   
The propagator $U(t) = e^{-\Omega t}$, and therefore $U^\dag = U^{-1}$ is unitary, since  $\Omega^\dag = -\Omega$.  Thus, \Eq{BlochEqMat} represents a rotation, albeit still most generally in complex space.  

\subsubsection{\textbf{Rotation in real space}}
\label{subsubsec:Liouville rotation}

In the case where the orthonormal basis states are Hermitian operators, the components of the density matrix (coherence vector or Stokes tensor, in this case) are real, and $\Omega$ is also a real antisymmetric matrix. The rotation of \Eq{BlochEqMat} is then a rotation in real, multidimensional Cartesian space, which is the generalization \cite{Fano, Hioe} to N-level systems  of the result obtained in \cite{Feynman}.  The quantum dynamics are thus fully classical in the additional dimensions exceeding 3D physical space.  However, classical rotations in more than three dimensions are only marginally less abstract than rotations in Hilbert space and Liouville space. 
More accessible insight can be obtained by mapping the real-valued quantum states to physical space as follows.

\subsubsection{\textbf{Exact mapping to classical coupled oscillators}}
\label{subsubsec:ExactSHO}

A textbook exercise for deriving the Larmor precession of a spin-1/2 in a static magnetic field $B_0$ differentiates the first-order derivative in Ehrenfest's theorem to obtain a harmonic oscillator equation for the time evolution of expectation values of the spin components transverse to $B_0$.  Similarly, 
differentiate \Eq{BlochEqMat} again to obtain
%--------------------------------------------------------------
     \begin{equation}
\bm{\ddot r} = \Omega^2\, \bm{r}.
\label{SHO}
     \end{equation}
%--------------------------------------------------------------
Since $\Omega^2$ is real, symmetric, and diagonalizable, the solution
is readily written in terms of the normal-mode solutions obtained by diagonalizing $\Omega^2$.  The eigenvalues of antihermitian $\Omega$ are pure imaginary, resulting in negative eigenvalues $-\omega_a^2$ $(a = 1, 2, 3, \ldots, n)$ for $n \times n$ matrix $\Omega^2$.  The $n$ distinct eigenvectors \ket{\omega_a} constitute a basis set satisfying the completeness relation $\sum_{\,a} \ket{\omega_a}\bra{\omega_a} = \Id$ (the identity element).  In this eigenbasis, \Eq{SHO} for each component $r_a = \bra{\omega_a} {r} \rangle$ becomes simply
%--------------------------------------------------------------
     \begin{equation}
\ddot r_a = -\omega_a^2\, r_a
\label{NormalModesSHO}
     \end{equation}
%--------------------------------------------------------------
with standard harmonic oscillator solution
%--------------------------------------------------------------
     \begin{equation}
r_a(t) = r_a(0) \cos\omega_a t + \frac{\dot r_a(0)}{\omega_a} \sin\omega_a t
\label{SolNormalModes}
     \end{equation}
%--------------------------------------------------------------
and $\bm{\dot r}$ dependent on $\bm{r}$ according to \Eq{BlochEqMat}, giving 
%--------------------------------------------------------------
     \begin{eqnarray}
\ket{r(t)} &=& \sum_{a = 1}^{n}\, \ket{\omega_a}\bra{\omega_a}\, r(t)\,\rangle  
                           \nonumber \\
       &=&  \sum_{a = 1}^{n}\, \ket{\omega_a}\bra{\omega_a}\,
           \left[
                  \cos\omega_a t + \Omega\frac{\sin\omega_a t}{\omega_a}\
           \right] \ket{r(0)}  \nonumber \\
       &=& U(t)\, \ket{r(0)}.
\label{U(t)SHO}
     \end{eqnarray}
%--------------------------------------------------------------

Representing the propagator in the eigenbasis where $\Omega$ is diagonal gives \Eq{NormalModesSHO}.  Using the representation for non-diagonal $\Omega$ and its eigenvectors to calculate $U(t)$ gives the physical displacements $r_i(t)$ for each of the $n$ oscillators.  This solution for $U(t)$ must be identical to the propagator given in \Eq{BlochSol}.  It is included primarily for consistency in the presentation, but also to emphasize the fundamental differential equation under consideration is first order and only requires specification of $\bm{r}(0)$ (or $\bm{r}$ evaluated at any other fixed time). 

\begin{figure}[ht]
\includegraphics[scale=.54]{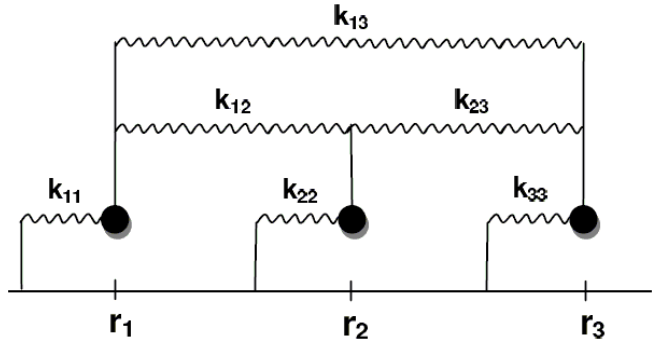}%[scale=1]{Fig1.eps}
\caption{Schematic of three masses at equilibrium positions $r_i = 0$ coupled with springs of stiffness $k_{i j}$.}
 \label{CoupledMasses}
\end{figure}

To complete the explicit identification of \Eq{SHO} with mechanical oscillators, consider equal masses, $m$, on a frictionless surface, with mass $m_i$ connected by spring of stiffness $k_{ij} = k_{ji}$ to mass $m_j$ ($i,j = 1,2,3,\dots,n$), as in \Fig{CoupledMasses} for an illustrative case $n=3$.  
%--------------------------------------------------------------
%------------------------------------------------------
The classical matrix $\Omega_\textrm{Cl}$ relating the displacement from equilibrium of the $i^{th}$ mass to its acceleration, as in \Eq{SHO}, is
%--------------------------------------------------------------
	\begin{equation}
(\Omega_\textrm{Cl})_{i j} = \frac{1}{m}  \left\{   
   \begin{array}{cl}
   k_{i j} & \quad i \neq j \\
    -\sum\limits_{l=1}^n |k_{i l}| & \quad i=j      
   \end{array}
\right.
\label{Omega_Cl}
	\end{equation}
%--------------------------------------------------------------

The above expression assumes reciprocal couplings $k_{ij} = k_{ji}$ and employs the absolute value in the sum to accommodate negative couplings.   A system of pendulums consisting of masses attached to rigid rods can be coupled negatively by attaching a spring to rod $i$ below the fulcrum of oscillation and to rod $j$ above the fulcrum.  Displacing mass $m_i$ to the right exerts a force on $m_j$ to the left, ie, the coupling $k_{ij} < 0$. A pendulum can be inverted with its mass above the fulcrum to implement $k_{ii} < 0$. Inverting transformers can be used to implement negative couplings in LC circuits. 

Setting $\Omega_\textrm{Cl} = \Omega^2$ gives the spring constants 
%--------------------------------------------------------------
     \begin{equation}
\frac{k_{i j}}{m} = \left\{   
   \begin{array}{cl}
   (\Omega^2)_{i j} & \quad i \neq j \\
    -(\,\Omega_{ii}^2 + \sum\limits_{l\neq i}^n |\Omega^2|_{i l}) & \quad i=j      
   \end{array}
\right.
\label{k_ij}
     \end{equation}
%--------------------------------------------------------------
in terms of the matrix $\Omega$ (squared) representing the quantum system, as derived from \Eq{Omega_ij}.  
There is thus a one-to-one mapping of the quantum states to the oscillator displacements embodied in both systems in $r_i(t)$. Given the initial states $r_i(0)$ of the system, the necessary $\dot r_i(0)$ follow from \Eq{BlochEqMat}.   

This mapping is very general.  It is not limited to particular values of the spin, numbers of interacting spins, specific forms of the commutation relations, or relative fractions of mixed and pure states comprising \dm. An $N\times N$ density matrix generates $N^2$ components in \Eq{rho(r_j)}, which requires $N^2$ oscillators. The static component of the identity element can be eliminated, and the structure of the Hamiltonian may generate evolution only in a smaller subspace of states, further reducing the number of required oscillators.

For pure states, the time-dependent elements $c_i(t)$ comprising the state vector can readily be obtained, if desired, from \dm\ reconstructed in matrix form using the $r_i(t)$ and \Eq{rho(r_j)}. Each resulting element $\rho_{ij}$ is equal to $c_i c^\ast_j$.  Assigning any one of the $c_i$ to the square root of $\rho_{ii}$ sets the arbitrary global phase of the pure-state elements.  In terms of this real $c_i$, the remaining $c_j$ are equal to $\rho_{ji}/c_i$.

\subsection{Schr\"odinger equation}
\label{subsec:Hilbert}
The solution 
%--------------------------------------------------------------
     \begin{equation}
\ket{\Psi(t)} = e^{-i\,H t}\, \ket{\Psi(0)}.
\label{SolTDSE}
     \end{equation}
%--------------------------------------------------------------
to the time-dependent Schr\"odinger equation
%--------------------------------------------------------------
     \begin{equation}
i\, \ket{\dot\Psi(t)} = H\, \ket{\Psi(t)}
\label{TDSE}
     \end{equation}
%--------------------------------------------------------------
represents a rotation of \ket{\Psi(0)} in Hilbert space, since $H$ is Hermitian and the propagator $U(t) = e^{-i\,H\, t}$ is unitary.  
Most generally, $H$ and the components $c_i$ of \ket{\Psi} in a chosen basis are complex, so there is no classical interpretation for the time evolution of the state.  

In \cite{Briggs2012b}, the authors derive a methodology for representing the complex $c_i$ in terms of the displacement and velocity of classical coupled oscillators for the special case of real, invertible \tHam.  These restrictions can be removed, as will be shown in what follows.

\subsubsection{\textbf{Rotation in real space}}
\label{subsubsec:Hilbert rotation}

The results of the preceding sections can be extended to the Schr\"odinger equation using only the quantum Hamiltonian, in contrast to the approaches in \cite{Dirac, Strocchi, Briggs2012b} which start with the real, classical Hamiltonian \tHam\ and relate it to the quantum Hamiltonian $H$.  The expression for \tHam\ in terms of $H$ then represents classical coupled harmonic oscillators only for real $H$ \cite{Briggs2012b}.

Instead, start with \Eq{TDSE} and represent the components of \ket{\Psi} as vector $\bm{c}$, giving
%--------------------------------------------------------------
     \begin{equation}
\bm{\dot c}\,(t) = -i H \bm{c}\,(t) .
\label{MatrixFormTDSE}
     \end{equation}
%--------------------------------------------------------------
Write complex $\bm{c} = \bm{q} + i\bm{p}$ and complex 
$H = Q + i P$.  For real $H$, $\bm{q}$ and $\bm{p}$ are conjugate variables \cite{Briggs2012b} but are not most generally so.  Equating the real and imaginary parts after performing the multiplications in \Eq{MatrixFormTDSE} recasts the Schr\"odinger equation as a real equation
%--------------------------------------------------------------
     \begin{eqnarray}
\left( \begin{array}{c}
           \bm{\dot q} \\ 
           \bm{\dot p}
       \end{array}
\right)  &=&  \left( \begin{array}{cc}
                         P & Q \\
                        -Q & P
                     \end{array}
             \right) 
             \left( \begin{array}{c}
                        \bm{q} \\
                        \bm{p}
                     \end{array}
              \right)  \nonumber \\
         &=& \Omega \left( \begin{array}{c}
                        \bm{q} \\
                        \bm{p}
                     \end{array}
              \right)
 \label{RealTDSE}
     \end{eqnarray}
%--------------------------------------------------------------
in the form of \Eq{BlochEqMat} 
with $\Omega^\dag = -\Omega$ again antihermetian, since Hermitian $H$ requires $Q^\dag = Q$ and $P^\dag = -P$.  The results and implications for real rotations then follow from $\S\,$\ref{subsubsec:Liouville rotation} and the discussion following \Eq{BlochEqMat}.

\subsubsection{\textbf{Exact mapping to classical coupled oscillators}}
\label{subsubsec:ExactHilbertSHO}

Similarly, differentiating \Eq{RealTDSE} gives
%--------------------------------------------------------------
     \begin{eqnarray}
\left( \begin{array}{c}
           \bm{\ddot q} \\ 
           \bm{\ddot p}
       \end{array}
\right) &=&   \left( \begin{array}{cc}
                       P & Q \\
                      -Q & P
                    \end{array}
             \right)^2 
             \left( \begin{array}{c}
                        \bm{q} \\
                        \bm{p}
                     \end{array}
              \right)  \nonumber \\
        &=& \left( \begin{array}{cc}
                       P^2 - Q^2 & P Q + Q P \\
                      -(P Q + Q P) & P^2 - Q^2
                    \end{array}
             \right)
             \left( \begin{array}{c}
                        \bm{q} \\
                        \bm{p}
                     \end{array}
              \right)  \nonumber \\
        &=& \left( \begin{array}{cc}
                      -\Re(H^2) & \Im(H^2) \\
                      -\Im(H^2) & -\Re(H^2)
                    \end{array}
             \right)
             \left( \begin{array}{c}
                        \bm{q} \\
                        \bm{p}
                     \end{array}
              \right)  \nonumber \\
        &=& \Omega^2 \left( \begin{array}{c}
                        \bm{q} \\
                        \bm{p}
                     \end{array}
              \right) 
 \label{RealSchSHO}
     \end{eqnarray}
%--------------------------------------------------------------
in the form of \Eq{SHO} for real symmetric (Hermitian) matrix $\Omega^2$ constructed from the real and imaginary parts of $H^2$.  The mapping of $q$ and $p$ to mechanical oscillators then follows from $\S$\ref{subsubsec:ExactSHO}.  

For complex $N\times N$ Hamiltonian, there are thus most generally $2 N$ mutually coupled oscillators.  There can be fewer oscillators and no mutual coupling between specific oscillators, depending on the structure of $H$.  The displacements $\bm{q}$ and $\bm{p}$ provide an exact one-to-one mapping to the real and imaginary components, respectively, of the quantum state $\ket{\Psi}$.
The state $\bm{c}(0) = \bm{q}(0) + i\,\bm{p}(0)$ uniquely determines the initial displacements, with the initial velocities then given by \Eq{RealTDSE}.  

The present treatment produces the results in \cite{Briggs2012b} in the case of real $H = Q$, $P=0$.  Note that $\bm{q}$ and $\bm{p}$ then evolve independently according to the same propagator, with no mechanical coupling between $q$ and $p$ oscillators.  The initial conditions are the only difference in the solutions.  Calculating $\bm{p} = H^{-1}\bm{\dot q}$ separately as in \cite{Briggs2012b} imposes an additional unnecessary restriction that $H$, already constrained in \cite{Briggs2012b} to be real, must be invertible (ie, no eigenvalues equal to zero). 

\subsubsection{\textbf{Extension to mixed states}}
A statistical mixture can not be represented in terms of a state $\ket{\Psi}$, but is written in terms of the probability $p_k$ for being in each of the possible states 
$\ket{\Psi_k}$ as a density matrix
%--------------------------------------------------------------
     \begin{equation}
\rho(t) = \sum_k p_k \ket{\Psi_k(t)} \bra{\Psi_k(t)}
\label{DensityMatrix}
     \end{equation}
%--------------------------------------------------------------
which evolves according to \Eq{Liouville}.  The results in \cite{Briggs2012b} and extensions in the previous section are limited to pure states evolving according to the Schr\"odinger equation. The methodology would appear to be inapplicable to mixed states.  However, the density matrix representing a given system is not unique. It is an average over the $\cal{N}$ constituents comprising a macroscopic system, which can be astronomically large, precluding an exact determination of the exact state of each of the $\cal{N}$ constituents.

But the identical density matrix can also be constructed from a completely specified set of $N \ll \cal{N}$ noninteracting pure states, with the $N^2$ elements of \dm\ determined from measurable macroscopic (average) properties of the system, such as energy or polarization.  In that case, both the weights $p_k$ and corresponding states are known exactly, so each $\ket{\Psi_k}$ can be used independently to construct a set of coupled oscillators representing the components $c^{(k)}_i(t)$ of $\ket{\Psi_k(t)}$.  In lieu of density matrix evolution 
$\rho(t) = U \rho(0) U^\dag$, the simpler and more efficient Schrodinger evolution $\ket{\Psi(t)} = U \ket{\Psi(0)}$ can be applied to each pure state $\ket{\Psi_k}$ comprising $\rho$ in \Eq{DensityMatrix}.  Subsequently, the weights $p_k$ can be used to calculate expectation values, measurement probabilities, or reconstruct the density matrix at later times $t$ if desired.

In addition, as shown in \cite{Skinner02}, at least one of the $\ket{\Psi_k}$ comprising the initial density matrix is redundant and can be removed from the calculation, since it provides a relatively uninteresting constant contribution to the system dynamics.  Choose one of the weights, for example, $p_1$.  The density matrix can be rewritten as $p_1$ times the identity element plus a ``pseudo'' density matrix constructed from the $\ket{\Psi_k}$ with weights $(p_k - p_1)$.  The term that is proportional to the identity element doesn't evolve in time under unitary transformations and can be ignored.  

Thus, the state \ket{\Psi_1} has been removed from the density matrix, along with any other \ket{\Psi_k} that had original weights $p_k = p_1$.  
In the general case of $m\geq 1$ degenerate statistical weights $p_k$, only $m < N$ of the \ket{\Psi_k} are required.  The number of oscillators is correspondingly reduced. Choosing the weight with the largest degeneracy provides the maximum reduction. The explicit contribution of each pure state to the system dynamics is readily apparent in utilizing this approach.  The density matrix at any given time is easily reconstructed as described in \cite{Skinner02}.

\section{Dissipative Systems}
\label{sec:DissSys}

The modifications necessary to model open systems as a set of damped oscillators can be found very generally, with minimum detail concerning the relaxation formalism. 
The Wangsness-Bloch equation expressing the evolution of the density operator in the presence of relaxation adds a relaxation operator term to \Eq{Liouville} that operates on the density matrix \cite{Fano, Wangsness-Bloch}.  Expanding \dm\ in a basis of orthonormal operators as in \Eq{rho(r_j)} gives the real equation
%--------------------------------------------------------------
     \begin{equation}
\bm{\dot r} = \Omega\, \bm{r} + R\, \bm{r} + F(\bm{r}_{eq}),
\label{RelaxMatEq}
     \end{equation}
%--------------------------------------------------------------
which can be transformed to a homogeneous equation as in \cite{Fano}.
The relaxation matrix $R$ must be symmetric for relaxation elements that act symmetrically between states of the system, with diagonal elements providing auto-relaxation rates and off-diagonal elements giving cross-relaxation.  The function $F$ adds a constant term incorporating the asymptotic decay of the system to the steady state in terms of the equilibrium state $\bm{r}_{eq}$.  Without this term, the solution decays to zero.

Differentiating again gives
%--------------------------------------------------------------
     \begin{eqnarray}
\bm{\ddot r} &=& (\Omega + R) \bm{\dot r}  \nonumber \\
             &=& \Omega\,[\,(\Omega + R) \bm{r} + F\,] + R \bm{\dot r},
\label{DampedSHO}
     \end{eqnarray}
%--------------------------------------------------------------
ie, a set of coupled oscillators with a velocity-dependent friction term and a constant applied force $\Omega F$.  A constant force in the harmonic oscillator equation merely shifts the origin of the coordinates.  However, the matrix multiplying $\bm{r}$, which determines the mechanical couplings as given in \Eq{k_ij}, is no longer symmetric due to the sum of antisymmetric $\Omega$ and symmetric $R$, resulting in non-reciprocal off-diagonal couplings.  

The precise role of non-reciprocal couplings in a classical model for quantum dissipative systems can be clarified by eliminating $\bm{\dot r}$ to obtain
%--------------------------------------------------------------
     \begin{eqnarray}
\bm{\ddot r} &=& (\Omega + R)^2 \bm{r} + (\Omega + R) F  \nonumber \\
             &=& \Gamma^2 \,\bm{r} + \Gamma F,
\label{AltDampedSHO}
     \end{eqnarray}
%--------------------------------------------------------------
a set of ideal (frictionless) coupled oscillators subjected to a constant applied force.  
In this case, however, the matrix $\Gamma^2$ is the sum of symmetric $\Omega^2 + R^2$ and antisymmetric $\Omega R + R \Omega$.  The former term corresponds to a set of undamped oscillators with symmetric couplings $k_{ij} = k_{ji}$ ($\S$\ref{subsubsec:ExactSHO}), modified in comparison to no relaxation by inclusion of $R^2$.  The normal-mode frequencies are also modified accordingly.  

Damping is provided by the antisymmetric part of $\Gamma^2$, which gives antisymmetric couplings $\gamma_{ij} = -\gamma_{ji}$ and total coupling $\mathcal{K}_{ij} = k_{ij} + \,\gamma_{ij}$.  The $\gamma_{ij}$ therefore represent couplings connected in parallel with the symmetric $k_{ij}$ and can be implemented, in principle, using magnetic materials and magnetic fields.  For a given positive $\gamma_{ij}$, a positive displacement of mass $m_j$ results in a positive force on $m_i$.  The resulting positive displacement of $m_i$ provides a negative force on $m_j$ due to $\gamma_{ji} < 0$ which opposes the original displacement of $m_j$ and damps the motion.  Stated differently, energy transferred from $m_j$ to $m_i$ is not reciprocally transferred back from $m_i$ to $m_j$, and the motion is quenched.  

The inhomogeneous term $\Gamma F$ in \Eq{AltDampedSHO} can, equivalently, be included in an augmented matrix $\tilde\Gamma$ formed by appending a column $\Gamma F$ to the right of $\Gamma^2$ and then adding a correspondingly expanded row of zeros at the bottom.  The vector $\bm{r}$ is then augmented by including a last element equal to one to obtain the equivalent homogeneous equation
%--------------------------------------------------------------
     \begin{equation}
\frac{d^2}{dt^2}\bm{\tilde r} = \tilde\Gamma\, \bm{\tilde r}.
\label{HomDampedSHO}
     \end{equation}
%--------------------------------------------------------------
The asymmetry of $\tilde\Gamma$ generates unphysical couplings that are not a problem theoretically but preclude a real, physical model.  However, $\tilde\Gamma$ is readily written as the sum of symmetric $\tilde\Gamma_S = (\tilde\Gamma + \tilde\Gamma^\dag)/2$ and antisymmetric
$\tilde\Gamma_A = (\tilde\Gamma - \tilde\Gamma^\dag)/2$, which determine the symmetric couplings $k_{ij}$ and antisymmetric couplings $\gamma_{ij}$, respectively, as above.

In comparison, the Schr\"odinger equation can only include relaxation in certain special cases amenable to complex energies in the Hamiltonian.  A typical application is the coupling between stable and unstable states and the resulting lifetimes of the states.  An example relating velocity-dependent damping of classical oscillators to a Schr\"odinger equation treatment was provided in \cite{Briggs2012b} in the weak-coupling limit.  However, neither this approximation nor the required complex energies can be applied more generally.  Even a simple two-level system with relaxation dynamics described by the Bloch equation cannot be addressed by the Schr\"odinger equation and requires the density matrix approach.

\section{Illustrative Examples}
Simple two-level systems are used as a prototype for implementing the quantum-classical mapping.  Although they are already known to be representable by classical rotations in three-dimensional physical space,
they provide sufficient detail to clarify the connection between (i) real rotations of N-level quantum states in $N^2-1$ dimensions and (ii) their mapping to classical oscillators in one-dimensional physical space.
Quantum evolution in (complex) Hilbert space is also readily compared to the evolution of the corresponding classical systems derived from Liouville and Hilbert approaches.
In what follows, vectors are written conveniently as rows in the text, but are to be understood as column vectors when used in matrix equations.

\subsection{Quantum solution}
In terms of real $\Delta_1$, $\Delta_2$ and complex $V = \omega_1 - i \omega_2$, the Hamiltonian for a general 2-level system can be written in terms of the Pauli matrices $\sigma_i\,(i = 1,2,3)$ and $\sigma_0 = \openone$ as
%--------------------------------------------------------------
     \begin{equation}
H = \left( \begin{array}{cc}
                         \Delta_1 & V \\
                         V^\ast & \Delta_2
                     \end{array}
             \right) 
    = \sum_{\alpha = 0}^3 \omega_\alpha \sigma_\alpha ,
\label{2-levelH}
     \end{equation}
%--------------------------------------------------------------
with $\omega_0 = 1/2(\Delta_1 + \Delta_2)$ and $\omega_3 = 1/2(\Delta_1 - \Delta_2)$.
The $\sigma_0$ term commutes with the other terms, so the propagator $U(t) = e^{-i H t}$ giving the Schr\"odinger equation solution as in \Eq{SolTDSE} 
is readily obtained in terms of $\omega_i\,\sigma_i$ ($i = 1,2,3$). 
The standard expansion of 
$e^{-i\,\bm{\omega}\cdot\bm{\sigma}\,t}$ using unit vector 
$\bm{\hat\omega} = \bm{\omega}/\omega$ gives
% $\cos\omega t - i\, \bm{{\hat\omega}\cdot\bm{\sigma}}\,\sin\omega t$.  
% The textbook result $e^{-i\, \bm{\hat\omega}\cdot\bm{\sigma}\, \omega t} = 
% \cos\omega t - i\, \bm{{\hat\omega}\cdot\bm{\sigma}}\,\sin\omega t$

% The $\sigma_0$ term commutes with the other terms, so the propagator $U(t) = e^{-i H t}$ giving the Schr\"odinger equation solution as in \Eq{SolTDSE} can be factored as 
% $e^{-i\,\omega_0\, t} e^{-i\, \bm{\omega}\cdot\bm{\sigma}\, t}$. The textbook expansion
% $e^{-i\,(\omega t) \bm{\hat\omega}\cdot\bm{\sigma}} = 
% \cos\omega t - i\, \bm{{\hat\omega}\cdot\bm{\sigma}}\,\sin\omega t$
%--------------------------------------------------------------
     \begin{eqnarray}
U(t) &=& e^{-i\omega_0 t} e^{ -i \bm{\omega}\cdot\bm{\sigma} }  \nonumber \\
     &=& e^{-i\omega_0 t}\, [\,\cos\omega t - 
                             i\, \bm{\hat\omega}\cdot\bm{\sigma}\,\sin\omega t\,]
                                \nonumber \\
     &=& e^{-i\omega_0 t} \left( \begin{array}{cc}
                                       a & b \\
                                       -b^\ast & a^\ast
                                    \end{array}
                             \right) .  
\label{2-levelU}
     \end{eqnarray} 
%--------------------------------------------------------------
The parameters $a,b$ obtained from expanding $\bm{\hat\omega}\cdot\bm{\sigma}$ and using the matrix forms for the $\sigma_i$ are
%--------------------------------------------------------------
     \begin{eqnarray}
a &=& \cos\omega t -i\,\hat{\omega}_3 \sin\omega t  \nonumber \\
b &=& -(\hat{\omega}_2 + i \hat{\omega}_1) \sin\omega t,
\label{C-K}
     \end{eqnarray}
%--------------------------------------------------------------
recognizable from classical mechanics as the Cayley-Klein parameters for a rotation by angle $2\omega t$ about $\bm{\hat{\omega}}$. 

Evolution of Schr\"odinger state $\ket{\Psi} \leftrightarrow (c_1,c_2)$ proceeds according to \Eq{SolTDSE}, with the corresponding density matrix states 
$\rho_{ij} = c_i c_j^\ast$ evolving according to \Eq{SolLiouville}.  The equivalent classical evolution is considered next.

\subsection{\textbf{Classical representation (Liouville equation)}}

Using the $\sigma_\alpha$ as the basis and inner product $\langle\,\sigma_\alpha\,|\,\sigma_\beta\,\rangle = 1/2 \text{Tr}\,(\sigma_\alpha\,\sigma_\beta) = \delta_{\alpha\beta}$ gives $\Omega_{0\alpha} = 0 = \Omega_{\alpha 0}$ according to \Eq{Omega_ij}.  The remaining $3\times 3$ matrix giving the non-zero couplings is easily determined using the commutation relations $[\sigma_i, \sigma_j] = 2\, i \epsilon_{ijk}\sigma_k$ written in terms of the usual Levi-Civita tensor $\epsilon_{ijk}$ (equal to $\pm 1$ for cyclic/anticyclic permutations of the indices $j,k,l = 1,2,3$ and zero otherwise) summed over repeated indices for slightly more concise notation. Then 
%--------------------------------------------------------------
     \begin{eqnarray}
\Omega_{ij} &=& \InnerProduct{[\,\sigma_i, \sigma_j\,]}{H}  \nonumber \\
&=& -2\epsilon_{ijk} \InnerProduct{\sigma_k}{\bm{\omega}\cdot\bm{\sigma}} \nonumber \\
&=& -2\,\omega_l\epsilon_{ijk} \InnerProduct{\sigma_k}{\sigma_l} \nonumber \\
&=& -2\,\omega_k\epsilon_{ijk} \nonumber \\
\Omega &=& 2 \left( \begin{array}{ccc}
                       0 & -\omega_3 & \omega_2 \\
                       \omega_3 & 0 & -\omega_1 \\
                       -\omega_2 & \omega_1 & 0
                     \end{array}
             \right).
\label{2-level Omega_ij}
     \end{eqnarray}
%--------------------------------------------------------------
The resulting equation of motion
%--------------------------------------------------------------
     \begin{equation}
\bm{\dot r} = \Omega\bm{r} = 2\,\bm{\omega}\times\bm{r}
\label{2-level rdot}
     \end{equation}
%--------------------------------------------------------------
represents a rotation of $\bm{r}$ about $\bm{\omega}$ at angular frequency $2\,\omega$, as expected from the FVH Theorem for arbitrary two-level systems.  The equivalence of quantum dynamics, in the case of 2-level systems, to a rotation in real, physical space cannot be generalized to arbitrary N-level systems. Representing quantum dynamics by a system of coupled harmonic oscillators in one physical dimension \textit{is} general for any value of $N$.

The coupling matrix is
%--------------------------------------------------------------
     \begin{equation}
\Omega^2 = 4\,\left( \begin{array}{ccc}
   -(\omega_2^2 + \omega_3^2) & \omega_1\omega_2 & \omega_1\omega_3 \\
    \omega_1\omega_2 & -(\omega_1^2 + \omega_3^2) & \omega_2\omega_3 \\
    \omega_1\omega_3 & \omega_2\omega_3 & -(\omega_1^2 + \omega_2^2)
                     \end{array}
             \right),
\label{2-level OmegaSq}
     \end{equation}
%--------------------------------------------------------------
giving three mutually coupled oscillators as in \Fig{CoupledMasses}. The couplings obtained from \Eq{k_ij} are
%--------------------------------------------------------------
     \begin{eqnarray}
k_{ij}/4 &=& \omega_i\,\omega_j \qquad i \neq j  \nonumber \\
k_{ii}/4 &=& \omega_j^2 + \omega_k^2 - \omega_i\,\omega_j - \omega_i\,\omega_k 
            \qquad i \neq j \neq k  \nonumber \\
       &=& \omega_j(\omega_j - \omega_i) + \omega_k(\omega_k - \omega_i).
\label{2-level k_ij}
     \end{eqnarray}
%--------------------------------------------------------------
For any possible ordering of the relative magnitudes of nonzero $\omega_i$, at least one of the $k_{ii}$ has to be negative.

\subsection{\textbf{Classical Representation (Schr\"odinger equation)}}

The matrix $\Omega$ leading to a solution for $(\bm{q}, \bm{p})$ as a rotation $e^{-\Omega t}$ of the initial state $(\bm{q}_0, \bm{p}_0)$ is comprised of the real and imaginary parts of $H$ as in \Eq{TDSE}, giving
% %--------------------------------------------------------------
%      \begin{equation}
% Q = \left( \begin{array}{cc}
%                          \Delta_1 & \omega_1 \\
%                          \omega_1 & \Delta_2
%                      \end{array}
%              \right) 
% \label{2-level Q}
%      \end{equation}
% %--------------------------------------------------------------
% and the imaginary part
% %--------------------------------------------------------------
%      \begin{equation}
% P = \left( \begin{array}{cc}
%                          0 & -\omega_2 \\
%                          \omega_2 & 0
%                      \end{array}
%              \right) 
% \label{2-level P}
%      \end{equation}
% %--------------------------------------------------------------
% as in \Eq{RealTDSE}, giving
%--------------------------------------------------------------
     \begin{equation}
\Omega = \left( \begin{array}{cccc}
   0 & -\omega_2 & \Delta_1 & \omega_1 \\
   \omega_2 & 0 & \omega_1 & \Delta_2 \\
  -\Delta_1 & -\omega_1 & 0 & -\omega_2 \\
  -\omega1 & -\Delta_2 & \omega_2 & 0
                     \end{array}
             \right)
\label{2-level Sch Omega}
     \end{equation}
%--------------------------------------------------------------
and coupling matrix
%--------------------------------------------------------------
\begin{widetext}
     \begin{equation}
\Omega^2 = \left( \begin{array}{cccc}
   -\Delta_1^2 -(\omega_1^2 + \omega_2^2) & 
        -\omega_1(\Delta_1 + \Delta_2) & 
             0 &  -\omega_2(\Delta_1 + \Delta_2) \\
   -\omega_1(\Delta_1 + \Delta_2) & 
       -\Delta_2^2 -(\omega_1^2 + \omega_2^2) & 
           \omega_2(\Delta_1 + \Delta_2)      & 0 \\
    0 & \omega_2(\Delta_1 + \Delta_2) & 
          -\Delta_1^2 -(\omega_1^2 + \omega_2^2) &
              -\omega_1(\Delta_1 + \Delta_2) \\
   -\omega_2(\Delta_1 + \Delta_2) &
       0 & -\omega_1(\Delta_1 + \Delta_2) &
            -\Delta_2^2 -(\omega_1^2 + \omega_2^2)
                     \end{array}
             \right).
\label{2-level Sch OmegaSq}
     \end{equation}
\end{widetext}
%--------------------------------------------------------------
Four coupled oscillators are needed to represent $(q_1,q_2,p_1,p_2) \equiv (r_1,r_2,r_3,r_4)$. The mutual couplings $k_{ij}$ ($i\neq j$) given by \Eq{k_ij} are the corresponding elements of $\Omega^2$.  The self-couplings for $i=1,2$ are
%--------------------------------------------------------------
     \begin{equation}
k_{ii} = \Delta_i^2 + \omega_1^2 + \omega_2^2 - 
          (\omega_1 + \omega_2)(\Delta_1 + \Delta_2),
\label{2-level H k_ii}
     \end{equation}
%--------------------------------------------------------------
with $k_{33}=k_{11}$ and $k_{44}=k_{22}$.  Negative couplings are required in general, except for the special case $\omega_2 = 0$ and $\omega_1 < 0$.

The operator $\Omega$ generates simultaneous rotations in the planes $r_i$-$r_j$ 
associated with the nonzero $\Omega_{ij}$.  The nonzero mutual couplings represent the noncommuting rotations in $\Omega$. One easily shows (see Appendix) that noncommuting rotations share a common coordinate axis in their respective rotation planes, such as $r_2$-$r_1$ and $r_1$-$r_3$.  Then $\Omega_{21}\Omega_{13} = (\Omega^2_{23})$ gives a nonzero mutual coupling $k_{23}$. A rotation in the $r_1$-$r_2$ plane \textit{does} commute with a rotation in the $r_3$-$r_4$ plane, so one expects the mapping from rotations to oscillators to generate at least one mutual coupling equal to zero. For the particular example here, the structure of $\Omega$ is such that $(\Omega^2_{13}) = 0 = (\Omega^2_{24})$, giving zero for $k_{13}$ and $k_{24}$. Mass 1 is not coupled to mass 3, and mass 2 is not coupled to mass 4.  

Thus, the rotation of a single vector or rotor in real 4-dimensional space can be viewed equivalently as 4 oscillators evolving in one dimension.  Similarly, for N-level systems, the evolution of the associated rotor in $N^2 -1$ real dimensions is equivalent to the evolution of $N^2-1$ oscillators in one physical dimension.

\subsection{Quantum dimer}

The quantum dimer example provided in \cite{Briggs2012b} is given by real $V = \omega_1$, $\omega_2=0$, and $\Delta_1 = \Delta_2 = \omega_0$, giving $\omega_3 = 0$. 

\subsubsection{Liouville approach}
The only nonzero elements of $\Omega$ in \Eq{2-level Omega_ij} are then $\Omega_{32} = 2V = -\Omega_{23}$, leading to diagonal entries $(\Omega^2)_{22} = (\Omega^2)_{33} = -4V^2$ as the only nonzero elements of $\Omega^2$ in \Eq{2-level k_ij}.  Thus, two uncoupled oscillators, each with natural frequency $2V$, represent this particular quantum system, with the initial conditions determining the specific details of the time evolotion.  

For $\Psi(t) = [c_1(t), c_2(t)]$ and $\Psi(0) = (1,0)$, one easily obtains $\bm{r}(0) = (0,0,1/2)$ using $r_i = 1/2 \text{Tr}\,(\sigma_i\, \rho)$, resulting in $\bm{\dot r}(0) = [0, -V, 0)]$ from \Eq{2-level Omega_ij}.  Then
%--------------------------------------------------------------
     \begin{equation}
\bm{r}(t) =\frac{1}{2}  \left( \begin{array}{c}
                         0 \\
                        -\sin 2V t \\
                         \cos 2V t
                               \end{array}
                        \right) \, ,
\label{DimerSol}
     \end{equation}
%--------------------------------------------------------------
which is the expected rotation about axis $\bm{\hat\omega} =\bm{\hat\omega_1}$ at angular frequency $2\omega_1 t = 2 V t$ given by \Eq{C-K}.  Since the two oscillators are out of phase by $90^\circ$, the system can actually be represented by a single oscillator---the position of one oscillator automatically gives the position of the other from a simple phasor diagram.

\subsubsection{Schr\"odinger approach}
Referring to the $2\times2$ block structure of $\Omega^2$ in \Eq{2-level Sch OmegaSq}, one finds off-diagonal blocks equal to zero, since they depend on the imaginary part of $V$.  The two remaining nonzero blocks on the diagonal generate independent evolution of $\bm{q}$ and $\bm{p}$.  The $\bm{q}$-block gives two coupled oscillators with mutual coupling $k_{12}=-2\omega_0 V$ and self-couplings $k_{ii} = (\omega_0 - V)^2$ from \Eq{2-level H k_ii}.  The $\bm{p}$-block gives identical couplings.  One can instead switch to a positive value for the mutual coupling, as in \cite{Briggs2012b}, since the normal-mode eigenvalues $-(\omega_0 \pm V)^2$ of $\Omega^2$ are only interchanged by changing the sign of $V$. This changes the sense of rotation generated by $H$ in Hilbert space and hence, by $\Omega$ in the real 4-dimensional space.  Using a positive coupling in this way captures the essential elements of the problem, but does not, strictly speaking, faithfully map the quantum system to the oscillator system.  More importantly, negative couplings cannot be avoided, most generally.  

With the definitions in $\S$\ref{subsubsec:Hilbert rotation}, the initial condition $\bm{c}(0) = (1,0)$ corresponds to $(\bm{q}_0, \bm{p}_0) = (1,0,0,0)$, which extracts the first column of $\Omega$ in the matrix multiplication of \Eq{RealTDSE} to give 
$(\bm{\dot q}_0, \bm{\dot p}_0) = (0,0,-\omega_0,-V)$.  The four oscillators must be set in motion with these initial positions and velocities for their displacments in a mechanical implementation to correspond to the evolution of $\ket{\Psi(t)} = [c_1(t), c_2(t)]$. 

However, the propagator $U(t)$ is readily obtained from \Eq{U(t)SHO} in terms of the eigenvectors $(1,1)$ and $(1,-1)$ for each $2\times2$ block on the diagonal, padded with zeros to give the appropriate four-element vector.  A solution for the motion requires only the initial displacements.  The given initial condition picks out the first column of $U(t)$ to reproduce the solution given in \cite{Briggs2012b}.  The Schr\"odinger equation requires four coupled oscillators for this particular example, in contrast to two uncoupled oscillators for the Liouville representation (equivalent to a single oscillator, since they are always $90^\circ$ out of phase).

\subsection{Symmetric unperturbed levels}

Consider $\Delta_1 = -\Delta_2 = \omega_3$, which arises in representing two unequal energy levels relative to the mean energy of the levels.  

\subsubsection{Liouville approach}

There are no nonzero elements of $\Omega^2$ derived from \Eq{2-level OmegaSq}.  The system is fully coupled, as illustrated in \Fig{CoupledMasses}, and represents the most general result for this approach.  The Schr\"odinger approach, discussed next, provides a simpler representation in this case.

\subsubsection{Schr\"odinger approach}

The matrix $\Omega^2$ of \Eq{2-level Sch OmegaSq} is now diagonal for any general complex perturbation $V$.  Four uncoupled oscillators, each with natural frequency $(\sum_i \omega_i^2)^{1/2}$, represent the system.  Specifying \ket{\Psi(0)} determines the initial conditions as discussed previously.  This is a very simple system, with each oscillator evolving independently.

The Liouville approach, by contrast, results in a relatively more complex system, albeit with one less oscillator.  Yet, for the dimer example, the Liouville implementation is much simpler than the Schr\"odinger implementation.  Which approach gives the simpler set of oscillators and couplings depends on the specific problem.

\subsection{Bloch equation with relaxation}

The solution of the Bloch equation for the time dependence of nuclear magnetization in a magnetic field is relatively simple for a field along the $z$-axis \cite{Bloch}.
As is well-known, the transverse magnetization precesses about the field at the Larmor frequency while decaying exponentially at a transverse relaxation rate $1/T_2$.  The longitudinal magnetization relaxes to the equilibrium magetization at a rate $1/T_1$. The mapping of this motion to a system of damped oscillators illustrates the procedure described in $\S$\ref{sec:DissSys}, as well as the role of non-reciprocal couplings in the model.  Physical insight that may have been overlooked in the past is readily apparent from classical textbook treatments of damped oscillations.

The inhomogeneous term $F$ in \Eq{RelaxMatEq} is $(0,0,M_0/T_1$, where $M_0$ is the equilibrium magnetization.  Vector $\bm{r}$ represents the nuclear magnetization.  Denoting $\omega_3$ as the Larmor frequency, the matrix $\Gamma = \Omega + R$ is
%--------------------------------------------------------------
     \begin{equation}
\Gamma = 
\left(
\begin{array}{ccc}
 -\frac{1}{T_2} & -\omega _3 & 0 \\
 \omega _3 & -\frac{1}{T_2} & 0 \\
 0 & 0 & -\frac{1}{T_1} \\
\end{array} 
\right) .
\label{BlochMat}
     \end{equation}
%--------------------------------------------------------------
As described earlier, appending a column $\Gamma F$ to the right of $\Gamma^2$ followed by a row of zeros at the bottom gives \Eq{HomDampedSHO} for the oscillator equation, with
%--------------------------------------------------------------
     \begin{equation}
\tilde\Gamma = 
\left(
\begin{array}{cccc}
 \frac{1}{T_2^2}-\omega _3^2 & \frac{2 \omega _3}{T_2} & 0 & 0 \\
 -\frac{2 \omega _3}{T_2} & \frac{1}{T_2^2}-\omega _3^2 & 0 & 0 \\
 0 & 0 & \frac{1}{T_1^2} & -\frac{M_0}{T_1^2} \\
 0 & 0 & 0 & 0 \\
\end{array}
\right)
\label{HomBlochMat}
     \end{equation}
%--------------------------------------------------------------
and $\tilde r_4 = 1$ augmenting $\bm{r}$ to represent a static component that incorporates the inhomogeneous term $\Gamma F$.  The necessary couplings are easily read from symmetric $\Gamma_S$ and antisymmetric $\Gamma_A$ that sum to give $\tilde\Gamma$:
%--------------------------------------------------------------
     \begin{eqnarray}
   \tilde\Gamma_S &=&
\left(
\begin{array}{cccc}
 \frac{1}{T_2^2}-\omega _3^2 & 0 & 0 & 0 \\
 0 & \frac{1}{T_2^2}-\omega _3^2 & 0 & 0 \\
 0 & 0 & \frac{1}{T_1^2} & -\frac{M_0}{2 T_1^2} \\
 0 & 0 & -\frac{M_0}{2 T_1^2} & 0 \\
\end{array}
\right)  \nonumber \\
   \tilde\Gamma_A &=&
\left(
\begin{array}{cccc}
 0 & \frac{2 \omega _3}{T_2} & 0 & 0 \\
 -\frac{2 \omega _3}{T_2} & 0 & 0 & 0 \\
 0 & 0 & 0 & -\frac{M_0}{2 T_1^2} \\
 0 & 0 & \frac{M_0}{2 T_1^2} & 0 \\
\end{array}
\right) .
\label{S-AS Mat}
     \end{eqnarray}
%--------------------------------------------------------------

A symmetric coupling $k_{34} = -M_0/(2 T_1^2)$ connected in parallel with antisymmetric (nonreciprocal) coupling $\gamma_{34} = -M_0/(2 T_1^2)$ provides the contribution to the final steady state magnetization $\tilde r_3$ through coupling to $\tilde r_4$.  The vanishing of $k_{43} + \gamma_{43}$ ensures there is no coupling from $\tilde r_3$ to change the static component $\tilde r_4$.  
Although there is no friction term in \Eq{HomDampedSHO}, the mechanism that damps $\tilde r_3$ is fairly transparent.  Since $k_{31}= 0 = k_{32}$, $\tilde r_3$ is only coupled to static $\tilde r_4$, which effectively shifts $\tilde r_3$ to $z = \tilde r_3 - M_0$, giving the equivalent equation $\ddot z = z/T_1^2$.  The self-coupling $k_{33}$ is the source of the imaginary natural frequency $i/T_1$. resulting in the standard damped solution $z(t) = z(0) e^{-t/T_1}$.

The mechanism for transverse relaxation is perhaps more interesting, given that the diagonal elements $\tilde\Gamma_{ii}$ ($i = 1,2)$ cannot be the source of the damping for the case $\omega_3 = 1/T_2$.  More generally, since 
$\tilde\Gamma_{11} = \tilde\Gamma_{22}$ for any value of $\omega_3$, the eigenvalues are $\tilde\Gamma_{11}$ plus the eigenvalues for the antisymmetric block, which are $\pm\, 2i\omega_3/T_2$. The normal mode frequencies, given by the square root of the eigenvalues, are $\omega_3 \pm i/T_2$.   The asymmetric coupling is the sole source of the imaginary frequency producing the required $e^{-t/T_2}$ decay of the transverse magnetization.  The displacements of $\tilde r_1$ and $\tilde r_2$ exert opposing forces on each other compared to the usual symmetric couplings, which produce oscillations, and the motion is damped.

\section{Conclusion}
General N-level quantum systems can be represented as an assembly of classical coupled oscillators.  The methodology provides the possibility for visual, mechanical insight into abstract quantum systems, as well as a metric for characterizing the interface between quantum and classical mechanics. There is a one-to-one correspondence between oscillator positions at time $t$ and quantum states of the system. The formalism presented includes both open (dissipative) and closed systems. For systems represented by a density matrix, the known evolution of states as rotations of a single coherence vector in a real (but unphysical) hyperspace of $N^2-1$ dimensions \cite{Fano} has been mapped here to the evolution of $N^2-1$ oscillators in one physical dimension.  The evolution of Schr\"odinger states has also been generalized here to real rotations in $2N$ dimensions, which can be mapped to $2N$ oscillators in one-dimensional physical space. The scaling of quantum systems to classical systems is therefore linear for Schr\"odinger states rather than quadratic for density matrix representations.

The results are applicable to time-independent Hamiltonians, which is sufficiently general for a great many cases of practical interest.  Time-dependent Hamiltonians (ie, driven systems) can be approximated to a chosen level of accuracy by a sequence of constant Hamiltonians over sufficiently short time steps.  Modeling a time-dependent quantum Hamiltonian as classical requires a change in the spring constants of the mechanical system at each time step, together with a reinitialization of the velocities derived from the new positions of the oscillators according to \Eq{BlochEqMat}.  By contrast, velocities in the natural dynamics of a system of coupled oscillators would not change instantaneously and discontinuously with a change in spring constants.   

\begin{acknowledgments}
The author acknowledges support from the National Science Foundation under grant CHE-1214006 and thanks N.~Gershenzon for assitance with the figure.
\end{acknowledgments}

% Create the reference section using BibTeX:
%\bibliography{basename of .bib file}

\end{document}